**Title:** Combining multi-site Magnetic Resonance Imaging with machine learning predicts survival in paediatric brain tumours.

**Authors:** James T. Grist PhD[1], Stephanie Withey PhD [1,2,3], Christopher Bennett PhD [1], Heather E. L. Rose PhD [1,2], Lesley MacPherson PhD [4], Adam Oates PhD [4], Stephen Powell MSc[1], Jan Novak PhD [2,5,6], Laurence Abernethy PhD [7], Barry Pizer PhD [8], Simon Bailey PhD [9], Steven C. Clifford PhD[10], Dipayan Mitra PhD [11], Theodoros N. Arvanitis PhD [1,2,12], Dorothee P. Auer PhD [13,14], Shivaram Avula FRCR [7], Richard Grundy PhD [15], Andrew C Peet PhD [1,2].

1. Institute of Cancer and Genomic Sciences, School of Medical and Dental Sciences, University of Birmingham, Birmingham, UK.
2. Oncology, Birmingham Women's and Children's NHS foundation trust, Birmingham, United Kingdom.
3. RRPPS, University Hospitals Birmingham NHS foundation trust, Birmingham, United Kingdom.
4. Radiology, Birmingham Women's and Children's NHS foundation trust, Birmingham, United Kingdom.
5. Psychology, College of Health and Life Sciences Aston University, Birmingham, United Kingdom.
6. Aston Neuroscience Institute, Aston University, Birmingham, United Kingdom.
7. Radiology, Alder Hey Children's NHS foundation trust, Liverpool, United Kingdom.
8. Oncology, Alder Hey Children's NHS foundation trust, Liverpool, United Kingdom.
9. Sir James Spence Institute of Child Health, Royal Victoria Infirmary, Newcastle upon Tyne, United Kingdom.
10. Wolfson Childhood Cancer Research Centre, Newcastle University Centre for Cancer, University of Newcastle, United Kingdom.
11. Neuroradiology, Royal Victoria Infirmary, Newcastle Upon Tyne, United Kingdom.
12. Institute of Digital Healthcare, WMG, University of Warwick, Coventry, United Kingdom.
13. Sir Peter Mansfield Imaging Centre, University of Nottingham Biomedical Research Centre, Nottingham, United Kingdom.

14. NIHR Nottingham Biomedical Research Centre, Nottingham, United Kingdom.
15. The Children's Brain Tumour Research Centre, University of Nottingham, Nottingham, United Kingdom.

Corresponding author: Professor Andrew Peet, a.peet@bham.ac.uk.

**Research in context**

An initial pubmed search was performed in December 2018 using the key words 'MRI', 'Survival', 'Paediatric brain tumour', 'Machine Learning', 'Diffusion', and 'Perfusion'. We found several studies using diffusion and perfusion weighted imaging to distinguish between high and low-grade brain tumours, and also between tumour types. Further evidence for assessing paediatric brain tumours with magnetic resonance spectroscopy were found, with two particularly focused on assessing survival with either glycine or tumour lipid production. However, no such studies assessing the use of diffusion and perfusion imaging in predicting survival in paediatric brain tumours was found. Furthermore, no studies utilising advanced machine learning algorithms to automatically assess and predict survival in paediatric brain tumours using medical imaging were found.

**Added value of this study**

This study provides a novel, non-invasive, framework for the early assessment of tumour risk, and therefore predicted survival, within the paediatric brain tumour population using multi-centre perfusion and diffusion weighted imaging. Combining imaging with advanced machine learning methods, elevated perfusion within a tumour (regardless of grade and histological subtype) was predictive of decreased survival probability, and when combined with diffusion imaging revealed a novel stratification of low-and high-risk tumours that bear dramatically altered survival characteristics. These sub-groups had a mix of both low- and high-grade tumours and contained a mixture of tumour types, and it was possible to automatically classify cases using supervised machine learning with high accuracy.

**Implications of all the available evidence**

The implications of our study are that the introduction of acquisition and interpretation of perfusion imaging at diagnosis plays a key role in the initial assessment of paediatric brain tumour risk. Furthermore, the assignment of new brain tumour cases to low or high-risk categories at diagnosis will allow for convectional therapy or recruitment to trails for experimental therapies to be undertaken, respectively.

## Abstract

### Background

Brain tumours represent the highest cause of mortality in the paediatric oncological population. Diagnosis is commonly performed with magnetic resonance imaging and spectroscopy. Survival biomarkers are challenging to identify due to the relatively low numbers of individual tumour types, especially for rare tumour types such as atypical rhabdoid tumours.

### Methods

69 children with biopsy-confirmed brain tumours were recruited into this study. All participants had both perfusion and diffusion weighted imaging performed at diagnosis. Data were processed using conventional methods, and a Bayesian survival analysis performed. Unsupervised and supervised machine learning were performed with the survival features, to determine novel sub-groups related to survival.
Sub-group analysis was undertaken to understand differences in imaging features, which pertain to survival.

### Findings

Survival analysis showed that a combination of diffusion and perfusion imaging were able to determine two novel sub-groups of brain tumours with different survival characteristics (p <0.01), which were subsequently classified with high accuracy (98%) by a neural network. Further analysis of high-grade tumours showed a marked difference in survival (p=0.029) between the two clusters with high risk and low risk imaging features.

### Interpretation

This study has developed a novel model of survival for paediatric brain tumours, with an implementation ready for integration into clinical practice. Results show that tumour perfusion plays a key role in determining survival in brain tumours and should be considered as a high priority for future imaging protocols.

### Funding

We would like to acknowledge funding from Cancer Research UK and EPSRC Cancer Imaging Programme, The Action Medical research and the Brain Tumour Charity (Children with Cancer 15/188), the Children's Cancer and Leukaemia Group (CCLG) in association with the MRC and Department of Health (England) (C7809/A10342), the Cancer Research UK and NIHR Experimental Cancer Medicine Centre Paediatric Network (C8232/A25261), the Medical Research Council – Health Data Research UK Substantive Site and Help Harry Help Others charity. Professor Peet is funded through an NIHR Research Professorship, NIHR-RP-R2-12-019. Stephen Powell gratefully acknowledges financial support from EPSRC through a studentship from the Physical Sciences for Health Centre for Doctoral Training (EP/L016346/1). Theodoros Arvanitis is partially funded by the MRC (HDR UK). We would also like to acknowledge the MR radiographers at Birmingham Children's Hospital, Alder Hey Children's Hospital, the Royal Victoria Infirmary in Newcastle and Nottingham Children's Hospital for scanning the patients in this study. We would also like to thank Selene Rowe at Nottingham University Hospitals NHS Trust for help with gaining MRI protocol information. Dr James Grist is funded by the Little Princess Trust (CCLGA 2017 15).

**Trail registration number:** 04/MRE04/41

**Introduction**

Brain tumours represent one of the most common causes of paediatric and adult oncological mortality. Particular challenges are faced in clinical paediatric oncology research due to the highly heterogeneous nature of paediatric tumours, combined with the relative rarity of the disease in the general population[1]. Despite this, multi-centre studies have allowed impressive advances to be made in the understanding of the major types of children's brain tumours and these are starting to change clinical practice[2,3]. The majority of studies have relied on analysis of tumour tissue; however, medical imaging is becoming increasingly able to probe tissue properties and has the advantage that measurements are made directly in vivo. This is particularly important for probing the tissue microenvironment since quantities such as perfusion cannot be readily determined in tissue samples. Imaging therefore has the potential to provide new biomarkers of prognosis which can be obtained early and throughout the patient journey.

Recently an increased understanding of paediatric brain tumour biology has enabled more accurate prognostication for individual patients. The findings have largely been based on molecular genetic markers identified in tissue. For example in medulloblastoma, biological subgrouping has shown that WNT subgroup tumours have an excellent prognosis whereas group 3 tumours and subsets of SHH tumours have an inferior outcome[4]. However, in even rarer tumours, such as atypical rhabdoid tumours (ATRT), or midline gliomas, where biopsy derived tissue is challenging to acquire, it is even more challenging to perform studies[5]. Therefore, biological studies have been more difficult to perform in meaningful numbers for many tumour types and the small biopsies taken may not provide a representative view of the tumour, particularly its microenvironment.

Medical imaging is an important diagnostic aid for brain tumours, since it is non-invasive and can include the whole tumour and surrounding tissue. It is also capable of probing the tumour microenvironment *in vivo*, improving our understanding of the *in vivo* neovascularisation and cellularity of the tumour, as well as surrounding cerebral tissue through perfusion and diffusion imaging, respectively[6,7]. However, as mentioned above for biological studies, recruiting large numbers of patients for imaging studies is challenging, and often requires large multi-centre trials to glean meaningful results. In spite of this, these non-invasive modalities represent highly attractive methods to derive crucial information surrounding the diagnosis and progression of tumours.

Diffusion imaging is available on every major commercial MRI scanner and is routinely used to assess brain tumors[8]. Apparent diffusion coefficient (ADC) maps represent the speed of water motion in the tissue and this correlated cellularity. Perfusion imaging is often acquired either with dynamic susceptibility contrast (DSC) or arterial spin labelling (ASL) techniques[9,10]. DSC imaging is undertaken through the introduction of an exogeneous contrast agent containing gadolinium, and the passage of this bolus through the cerebral vasculature is rapidly imaged and post-processed to form quantitative cerebral blood volume and flow maps[11].

Studies have shown that diffusion and perfusion imaging are able to discriminate between paediatric tumour types *in vivo,* with high cellularity and perfusion in high grade tumours, and vice versa for low-grade[12,13]. This data, in turn, has informed survival analysis models using traditional methods such as Cox-regression to derive significant covariates from imaging data[14]. In particular, ADC mean, elevated cerebral blood flow, and image derived texture parameters have been found to be significant factors in long-term paediatric brain tumour survival[7,15,16].

In this study we have taken a novel approach to the understanding of risk and survival in paediatric brain tumours. We have combined a cohort of patients with multiple tumour types, including both common and rare tumours, and grades from multiple clinical centres. All participants had both perfusion and diffusion imaging, we have employed both supervised and unsupervised machine learning to determine key imaging derived risk factors to further our understanding and prediction of survival in the paediatric neurooncological population.

**Methods**

**Patient recruitment and imaging**

69 participants with suspected brain tumours (medulloblastoma (N = 17), pilocytic astrocytoma (N = 22), ependymoma (considered high grade, N = 10), other tumours (N = 20) are found in supplementary document 1. They were recruited from four clinical sites in the United Kingdom (Ethics reference: 04/MRE04/41, Birmingham Children's Hospital, Newcastle Royal Victoria Infirmary, Queen's Medical Centre, Alder Hey Children's Hospital, Liverpool). Participants underwent MRI, protocol discussed below, before invasive biopsy to confirm diagnosis. The median follow-up time for the cohort was 4.4 years. Tumours were assigned to high- (3&4) and low-grade (1&2) groups, with full cohort details found in supplementary document 1.

The imaging protocol for all participants was performed either at 3 or 1.5T and included standard anatomical imaging ($T_1$-weighted pre- and post- contrast and $T_2$-weighted) as well as diffusion and dynamic susceptibility contrast imaging covering the tumour volume (imaging sequence details found in supplementary Table 1). Additional clinical data (age at diagnosis and gender) were also collected for analysis.

**Image post-processing and analysis**

ADC maps were calculated from diffusion weighted imaging using a linear fit between the two b-value images in Matlab (The Mathworks, MA, 2018a). DSC time-course data were processed using conventional methods to provide uncorrected cerebral blood volume (uCBV) maps, with a leakage correction undertaken to produce corrected cerebral blood volume (cCBV) and K2 maps[17].
$T_2$- weighted imaging and ADC maps were registered to the first DSC volume with SPM12 (UCL). Regions of interest segmenting the tumour volume were drawn on the $T_2$ weighted imaging[18].

Image analysis was performed in Matlab (2018b, The Mathworks, MA), with the image mean, standard deviation, skewness, and kurtosis were calculated on a volume by volume basis for ADC and UCBV/CCBV/K2 maps for regions of interest and the whole brain as previously described[18]. Tumour volume (cm$^3$) was calculated from the T$_2$ ROI masks drawn by S.W[18]. Regions of interest were also drawn in normal appearing deep grey and white matter for each participant to calculate average diffusion and perfusion measures in normal appearing tissue by J.G. Medulloblastoma Chang stage was derived from radiological reports.

**Histological and Genetic analysis**

Histological (including MiB1, Ki67, Glial fibrillary acidic protein (GFAP), INI-1, Isocitrate dehydrogenase-1 (IDH-1), Neuron specific enolase (NSE), S-100, BAF47, BRAF fusions, P53) and genetic data (MYC status and medulloblastoma sub-type), where available, were collected from local sites and are found in supplementary document 1. Medulloblastomas were analysed for histological type, subgroup, and MYC and MYCN amplification status were determined by protocols established at Newcastle University[19–21]. Medulloblastoma histology was centrally reviewed at the Royal Victoria Infirmary. Data are summarised in supplementary document 1.

**Statistical analysis**

All statistical analyses were performed in R (3.6.1) with significance defined at $p<0.05$, and Bonferroni correction for multiple comparisons used where appropriate.
The data processing pipeline used in this study is summarised in Figure 1.

**Univariate statistical analysis**

Data normality was assessed using a Shapiro-Wilk test. Subsequently, differences in clinical and imaging features between high- and low-grade tumours were assessed using unpaired t-tests or Mann-Whitney U tests, where appropriate. Area under the ROC curve (AUC) values were calculated for each imaging feature for high/low grade discrimination. Differences in high-/low- risk (defined below) participants were assessed using unpaired two-tailed t-tests or Mann-Whitney U test, depending on data normality.

After unsupervised clustering (described below), further Mann-Whitney U tests were performed to assess for differences in imaging features between low grade tumours in low and high-risk categories, and between alive high-grade tumours in low- and high-risk categories.

**Survival and correlation analysis**

Univariate Cox-regression was performed with each individual imaging feature, clinical data, and tumour grade used to assess survival hazard coefficients. Tumour grade and type were not used in the analysis detailed below.

Iterative Bayesian survival analysis was undertaken using the iterative BMAsurv package in R using 5-fold stratified cross validation to determine the posterior probabilities and coefficients of the top 5 imaging features that best describe the survival data[22]. Iterative analysis including up to 15 data features in combination at any one time.

**Unsupervised and supervised machine learning**

K means clustering was performed with the imaging features from Bayesian survival analysis, with the optimal number of clusters determined from the largest average silhouette width. Groups were clustered into high and low risk groups, and subsequently used for further Kaplan-Meier analysis to assess for differences in survival between clusters.

Supervised machine learning using the aforementioned Bayesian features was used to predict high/low risk groupings using the Orange toolbox (Orange) in Python (3.6), with Random Forest, a single layer Neural Network, and a support vector machine used. Validation of classifiers was performed using 10-fold stratified cross-validation.

Clinical and imaging data were subset into Whole Brain (WB), and Region of Interest (ROI) features, and tumour volume and used for supervised learning. Principal component analysis was used to reduce data dimensionality with 95% of data variance or N-1 (where N is the size of the smallest group) used. The top 5 Bayesian features were also used as input into the classifiers, with no further principal component analysis performed. Classifier performance was determined from the classifier accuracy (% correctly classified cases) and F-statistic.

## Results

A total of 69patients were analysed in this study with 33 imaging features, including tumour volume, derived per patient. Example tumour anatomical, diffusion, and perfusion imaging can be seen in Figure 2. The survival curve for the whole cohort is seen in Figure 3A, showing 75% overall survival.

### Diffusion and perfusion imaging can detect differences between tumour grade

Univariate statistical analysis showed significant differences in both whole brain and ROI imaging features between all high and low-grade tumours (feature with highest AUC = ADC mean (0.82) range: 0.63-0.82) full results detailed in supplementary Table 2. Grey and white matter imaging results are detailed in supplementary Tables 3.

### Perfusion imaging is plays a key role in assessing survival in paediatric brain tumours

Whole cohort univariate cox regression revealed a number of imaging features with significantly elevated hazard ratios (HR), for example Uncorrected CBV ROI mean (HR = 3.1, Confidence Intervals (CI) = 1.5-6.6, $p = 0.003$), full results detailed in Table 1A.
Bayesian analysis revealed the 5 most likely features to predict survival (probability that the feature coefficient is greater than 0, posterior coefficient) to be uCBV ROI mean (96%, 0.85), K2 ROI mean (39%, -0.17), uCBV whole brain mean (40%, 0.3), tumour volume (27%, 0.05), and ADC ROI kurtosis (20%, 0.02). Full results detailed in Table 1B.

### Unsupervised clustering detects distinct groups with significantly different survival and imaging characteristics

Using the Bayesian imaging features, k means clustering revealed two distinct clusters, shown in Figure 3B, which when combined with Kaplan-Meier analysis revealed a significant difference between a high and low risk population (see Figure 3C, $p = 0.0015$ – overall survival for high and low risk = 55% & 90%, respectively). Cox regression revealed an elevated Hazard Ratio (HR = 5.6, confidence intervals = 1.6-20.1, $p < 0.001$) for the high-risk cluster, relative to the low risk cluster.
Further univariate analysis of each cluster showed significant differences in a number of imaging features, for example elevated ADC kurtosis in high vs low risk clusters (10.1 ± 5.3 vs 4.3 ± 1.8, p<0.001, respectively). A combination of both high- and low-grade tumours were found in both clusters, all other results detailed in Table 2.

**Supervised machine learning can be used to distinguish between high/low risk clusters**
Supervised machine learning using imaging features showed that the Bayesian features combined with a single layer neural network, after stratified 10-fold cross validation, provided the most accurate classification of high- and low-risk patients (accuracy = 98%, F-statistic = 0.98).

**There is a distinct difference in survival between high- and low- risk high-grade tumours**
Further Kaplan-Meier analysis of clustered high-grade tumours revealed a significant difference in survival (p < 0.05) with a hazard ratio of 7 (0.9-53 lower and upper bounds, respectively). The Kaplan-Meier curves for high grade tumours in both clusters can be seen in supplementary figure 1. Further to this, it is noted that there are a number of children alive at study end with high-risk tumours and currently limited follow-up, for example a Choroid Plexus Carcinoma with a current follow-up of 1 year and a national average 5-year survival rate of 26%[23] and a medulloblastoma with less than 3-year follow-up and M3 Chang stage. There was no detectable difference in survival between the high- and low- risk groups within the low grade tumours (p>0.05). Imaging of example cases by risk and grade given in Figure 4.

Qualitative sub-group analysis of histology and genetics between low- and high-risk medulloblastomas revealed no significant differences between MYC amplification or groupings. The high-risk cluster exhibited a trend toward having a larger number of high Chang stage Medulloblastomas (M3 = 6, M2 = 4, M1 = 3) in comparison to the low risk cluster (M2 = 1, M1 = 1, M0 = 2) – data shown in supplementary document 1.

**Discussion**

This study has shown the power of combining diffusion and perfusion imaging with machine learning to predict survival risk in a mixed cohort of paediatric brain tumours. A small handful of studies have previously looked at assessing survival with one of the aforementioned imaging techniques[24–26]; however, here we have shown the utility of combined diffusion-perfusion measures to provide advanced modelling of survival. The univariate results assessing low/high grade suggested a number of key diffusion and perfusion features for the discrimination between groups, however most had a poor AUC. Therefore, this represented an ideal situation for the use of machine learning to combine these features to provide highly accurate classifiers to solve this challenge. Interestingly, the majority of parameters predicting survival were from the perfusion imaging which is not currently part of routine clinical practice in many centres. DWI has become a standard method for investigating childhood brain tumours and low ADC is seen as being a marker for higher cellularity and grade which would be associated with poorer survival. The current study substantiates this but shows that DSC-MRI may be an even better modality for predicting survival. The importance of the vessel leakiness parameter K in survival prediction also implies that DSC-MRI may have advantages in survival prediction beyond that available from methods which do not include the injection of contrast agent such as ASL. Furthermore, clustering demonstrated a reasonable separation of high (M1 to M3) from low (M0) Chang stage tumours suggesting that these imaging features identify some properties in the primary tumour which are associated with metastatic potential.

The unsupervised machine learning identified two groups of tumours which did not correspond to any obvious non-imaging tumour characteristics. The credibility of these groups as being distinct entities was substantiated by the high accuracy (98%) with which the tumours could be assigned to the correct group by a supervised learner. A number of patients in the high-risk cluster were still alive at the study end although some of these, including those from known poor prognostic groups had short follow-up times. Further analysis showed that a number of surviving high-risk low-grade tumours had imaging features similar to high grade tumours (such as elevated ADC kurtosis and CBV) and were significantly different to low-risk low grade tumours. It will be interesting to ascertain the clinical course of these tumours over longer periods of follow-up.

A particular strength of this work is that imaging features with clinical data provide a non-invasive tool that can assess risk early in the patient journey. Indeed, the use of a supervised classifier to predict risk category allows for the prospective integration of this model into a clinical decision support system – whereby radiological analysis of a small number of imaging features can rapidly identify patients that should be considered for inclusion into clinical trials for prospective evaluation and subsequent stratification. The use of *in vivo* imaging also has the advantage that it provides information that cannot be found from analysis of resected tissue, perfusion in particular is inherently an in vivo property.
A further strength of this study is the use of multi-site, multi-scanner data – providing reassurance that the results are robust to the natural variability that occurs in protocols and scanners within clinical practice. Using multiple centres also provided a more statistically powerful study from which clinically relevant results could be obtained.

The imaging modalities used in this study are widely available and so data acquisition should be readily achieved in routine clinical practice. The image processing and classification should be made available by integration into a clinical decision support tool which are increasingly being developed[27]. Indeed, the results shown above show that it is possible to stratify patients into high and low risk groups with a trained supervised neural network, therefore enabling further real-time decisions to be made with regards to appropriate clinical management and inclusion into research trials for novel therapies to aid those with the current worst prognosis.
With the current uncertainty surrounding the use of Gadolinium in clinical practice, and the inability to be used in patients with impaired renal function, future work will include the addition of arterial spin labelling (ASL), a technique to estimate perfusion without the introduction of exogenous contrast agents, as data from this technique has been shown to correlate well with DSC cerebral blood volume[28,29]. However, information on vessel leakiness will not be available from ASL.

In conclusion, this work has demonstrated a highly novel clinical application of advanced survival modelling and machine learning to non-invasively stratify patients for according to risk. Both diffusion and perfusion were found to be important in determining risk with perfusion contributing to a greater extent emphasising the importance of acquiring perfusion imaging. This work represents an important step forward in the use of machine learning to predict survival and paves the way for further clinical studies focusing on the successful identification and treatment of high-risk children with brain tumours.

**Tables**

**Table 1** – Cox Regression (A) and Bayesian survival (B) results.

A

| Feature | Beta | Hazard Ratio | 95% Confidence Interval | Significance |
|---|---|---|---|---|
| CBV ROI Uncorrected Mean | 1.13 | 3.1 | 1.5-6.6 | p = 0.003 |
| CBV Uncorrected Standard Deviation | -1.12 | 0.33 | 0.11-0.99 | p = 0.05 |
| K2 ROI mean | -2.02 | 0.13 | 0.03-0.63 | p = 0.011 |
| CBV Uncorrected Whole Brain Mean | 1.12 | 3.02 | 1.06-8.91 | p = 0.04 |

B

| Feature | Probability (%) | Posterior coefficient |
|---|---|---|
| Tumor volume | 27 | 0.05 |
| CBV ROI uncorrected mean | 96 | 0.85 |
| K2 ROI mean | 39 | -0.17 |
| ADC ROI Kurtosis | 20 | 0.02 |
| CBV Uncorrected WB Mean | 40 | 0.3 |

**Table 2** – Low and high-risk cluster group features

| Feature | Low risk | High risk | Significance |
|---|---|---|---|
| Male: Female | 16:19 | 19:15 | N/A |
| Low: High Grade | 23:12 | 7:27 | N/A |
| Censored: Events | 32:3 | 20:14 | N/A |
| Tumor Volume ($cm^3$) | 2.3 ± 2.8 | 5.6 ± 7.0 | p = 0.015 |
| ROI ADC Kurtosis | 4.3 ± 1.8 | 10.1 ± 5.3 | p < 0.001 |
| ROI ADC Skewness | 0.1 ± 1.0 | 2.1 ± 1.0 | p < 0.001 |
| ROI K2 mean ($min^{-1}$) | 0.0018 ± 0.0027 | -0.005 ± 0.002 | p < 0.001 |
| ROI CBV Uncorrected standard deviation (mL $100g^{-1}$ $min^{-1}$) | 1.44 ± 0.74 | 0.88 ± 0.42 | p < 0.001 |
| K2 whole brain standard deviation ($min^{-1}$) | 0.03 ± 0.02 | 0.019 ± 0.008 | p = 0.007 |
| CBV Corrected whole brain mean (mL $100g^{-1}$ $min^{-1}$) | 1.14 ± 0.27 | 1.29 ± 0.26 | p = 0.02 |

**Figure captions**

**Figure 1** – Data processing pipeline used in this study.

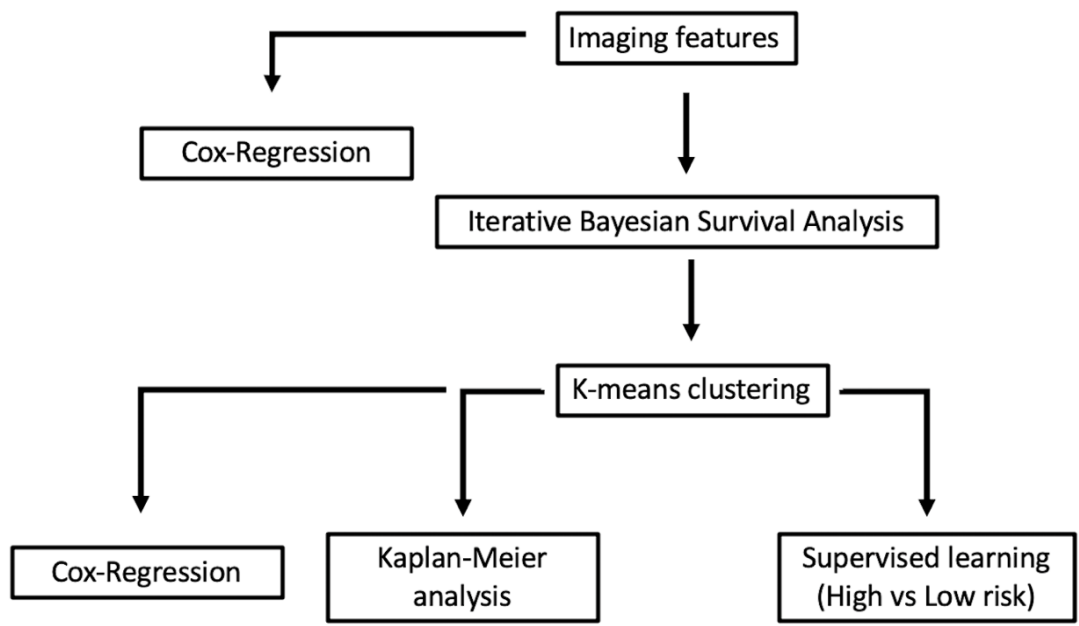

**Figure 2** – Example T$_2$ weighted, diffusion, and perfusion imaging of Ependymoma (A, B, and C respectively), Pilocytic astrocytoma (D, E, and F, respectively), choroid plexus carcinoma (F, G, and H respectively) and a Glioblastoma (I, J, and K, respectively).

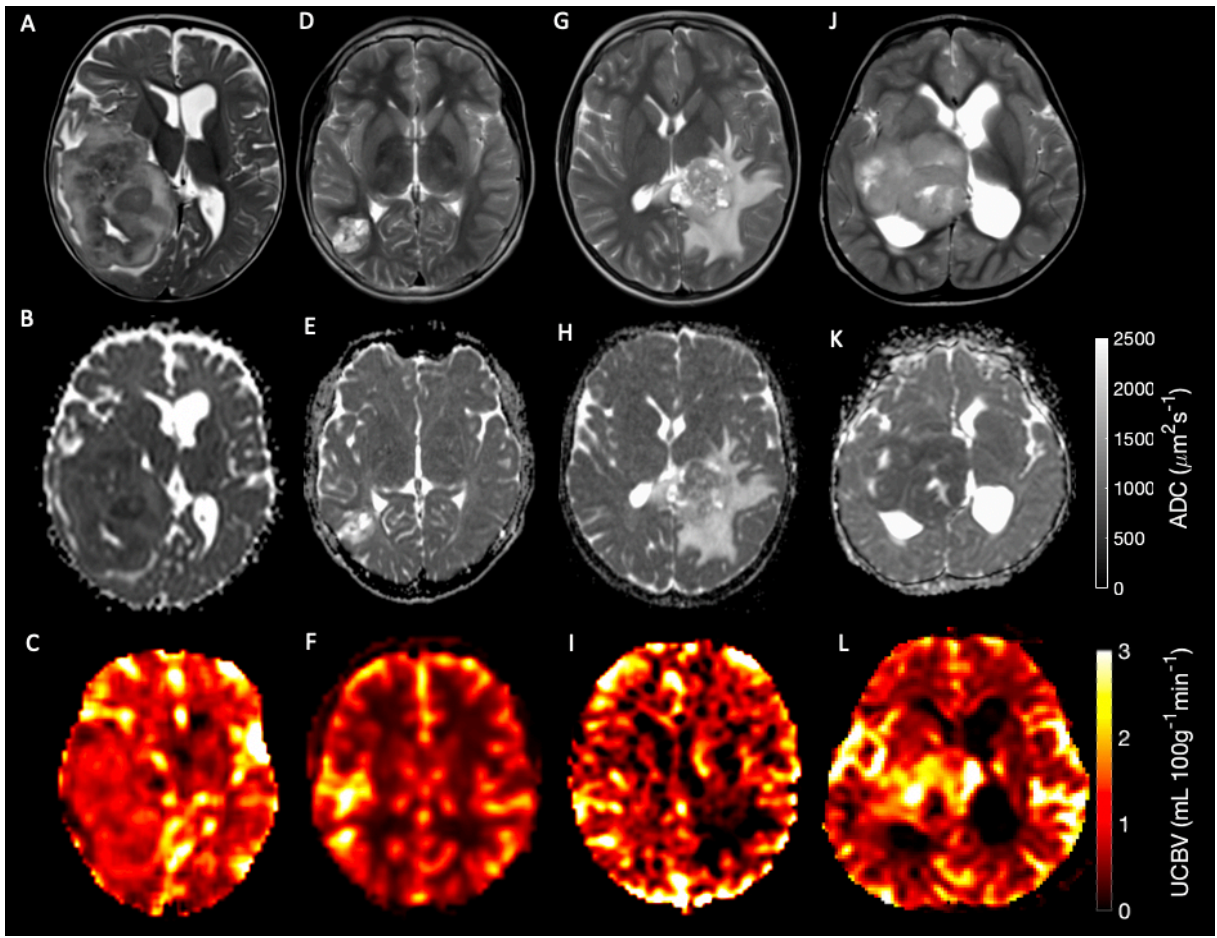

Figure 3 – (A) Overall survival curve for the cohort, (B) K Means clustering survival results showing two distinct clusters, (C) Kaplan-Meier curve for the two clusters showing a significant difference in survival. 1 = High risk, 2 = Low risk

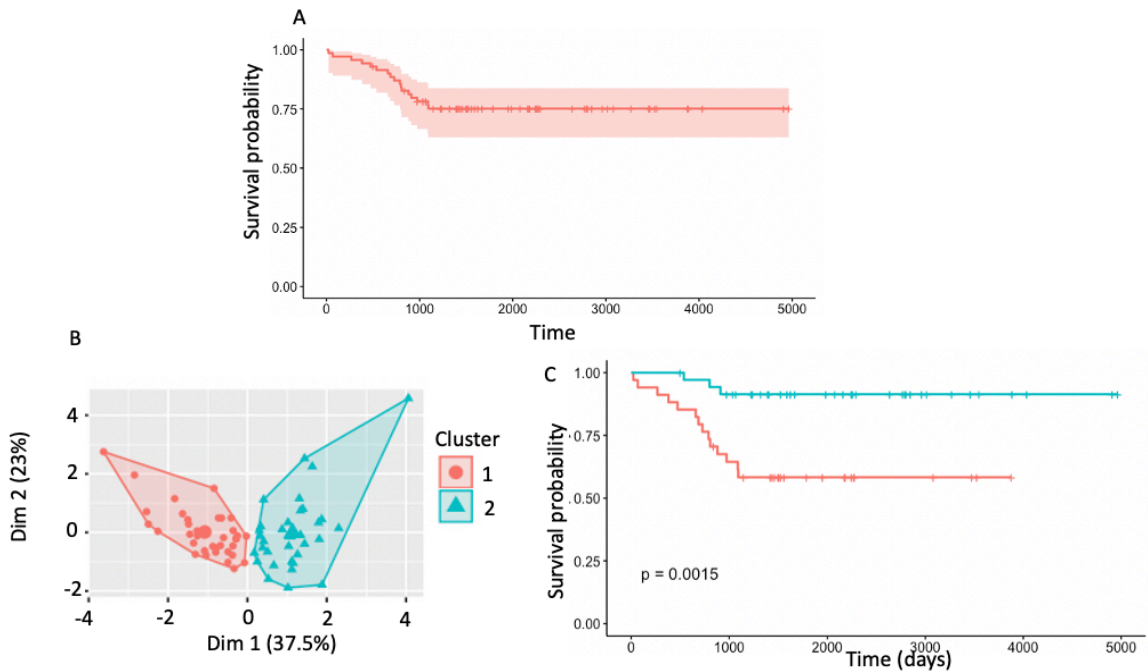

Figure 4 – Example high and low risk, high and low-grade tumors. (A T1 post contrast & B ADC map) high risk and (C T1 post contrast & D ADC map) low risk Pilocytic astrocytoma, respectively showing elevated ADC skew and kurtosis in the tumor region. (E & F) high risk and (G and H) low risk medulloblastomas, respectively, showing increased ADC kurtosis.

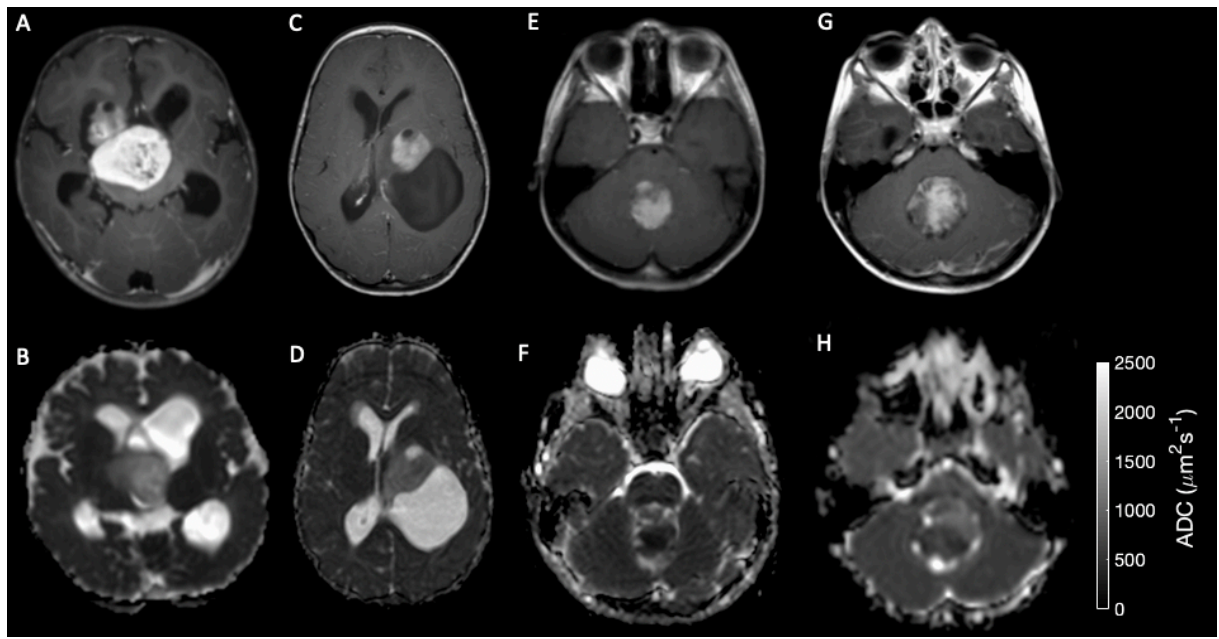

Supplementary figures and tables

**Table S1 - Imaging parameters used in this study.**

| Imaging sequence | Voxel volume (mm³) | Flip angle (degrees) | Repetition time (ms) | Echo time (ms) | B-values | Temporal resolution (s) |
|---|---|---|---|---|---|---|
| $T_2$-weighted | 1.6-2.6 | 90 | 3000-5230 | 80-108 | N/A | N/A |
| Diffusion weighted imaging (DWI) | 25-45 | 90 | 2600-5000 | 48-68 | 0, 800-1000 | N/A |
| DSC imaging (Echo planar imaging) | 11-45 | 20-75 | 582-2343 | 18.4-40 | N/A | 0.6-2.3 |
| DSC imaging (sPRESTO) | 18-22 | 7 | 15.5-17.2 | 23.5-25.2 | N/A | 1.2-1.6 |

**Table S2** - Differences in imaging features between high and low-grade tumors. (A) shows region of interest and (B) whole brain results. AUC = area under the curve.

A

| ROI Feature | Low grade | High grade | Significance | AUC |
|---|---|---|---|---|
| ADC mean ($\mu m^2/s$) | 1466 ± 437 | 962 ± 239 | 0.001 | 0.82 |
| ADC kurtosis | 5 ± 3 | 9 ± 5 | 0.001 | 0.74 |
| ADC skewness | 0.9 ± 1 | 2 ± 1 | 0.001 | 0.76 |
| CBV corrected mean (mL $100g^{-1}$ $min^{-1}$) | 1.95 ± 1.40 | 2.78 ± 1.93 | 0.04 | 0.67 |

B

| WB Feature | Low grade | High grade | Significance | AUC |
|---|---|---|---|---|
| ADC mean ($\mu m^2/s$) | 913 ± 302 | 713 ± 270 | 0.006 | 0.74 |
| ADC kurtosis | 5.2 ± 1.5 | 6.9 ± 2.2 | 0.009 | 0.71 |
| K2 standard deviation ($min^{-1}$) | 0.029 ± 0.018 | 0.019 ± 0.009 | 0.003 | 0.74 |
| CBV corrected mean (mL $100g^{-1}$ $min^{-1}$) | 1.13 ± 0.30 | 1.27 ± 0.22 | 0.03 | 0.63 |
| CBV corrected standard deviation (mL $100g^{-1}$ $min^{-1}$) | 1.02 ± 0.36 | 1.20 ± 0.36 | 0.04 | 0.67 |

**Table S3** - Average gray and white matter diffusion and perfusion values for the brain tumor cohort

| Feature | Gray Matter | White Matter | Significance |
|---|---|---|---|
| ADC mean ($\mu m^2/s$) | 912 ± 106 | 788 ± 64 | 0.003 |
| CBV uncorrected mean (mL $100g^{-1}$ $min^{-1}$) | 1.98 ± 0.26 | 1.04 ± 0.48 | 0.009 |
| K2 mean ($min^{-1}$) | 0.005 ± 0.002 | -0.001 ± 0.001 | 0.005 |
| CBV corrected mean (mL $100g^{-1}$ $min^{-1}$) | 2.47 ± 0.36 | 1.01 ± 0.43 | 0.008 |

**Figure S1 - Kaplan-Meier curves for high-grade low-risk (red) and high-risk (green) patients showing a significant difference in survival from imaging at diagnosis.**

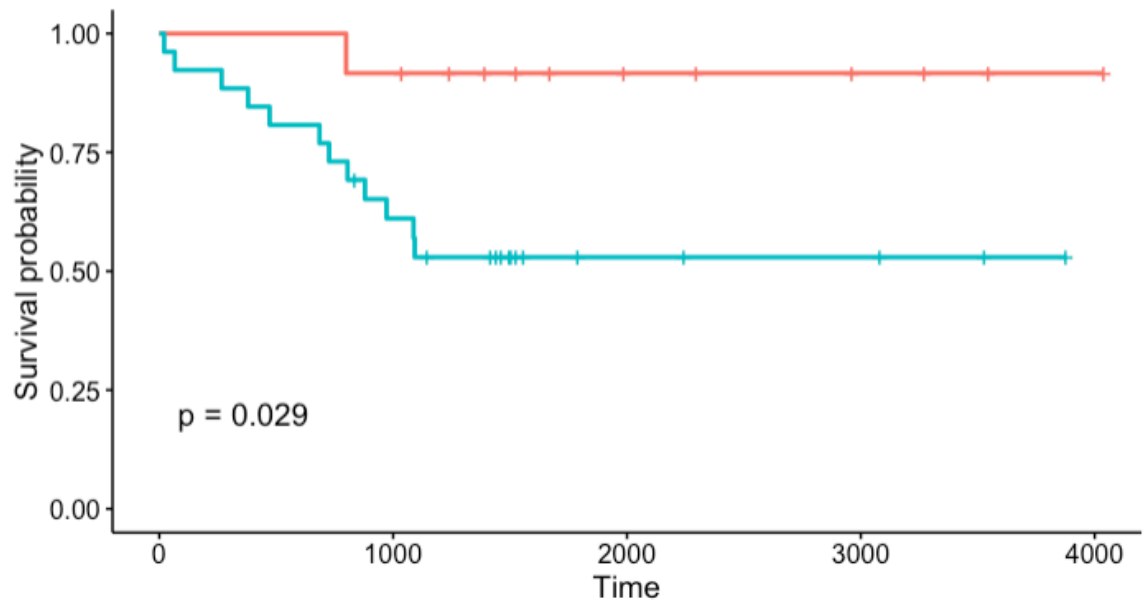